\documentclass{osa-article}
\usepackage{subfigure}
\journal{oe}

\begin{document}

\title{Ultranarrow spectral line of the radiation in double qubit-cavity ultrastrong coupling system}

\author{Teng Zhao,\authormark{1,2} Shao-ping Wu,\authormark{1} Guo-qing Yang,\authormark{3} Guang-ming Huang,\authormark{1} and Gao-xiang Li\authormark{1,*}}

\address{\authormark{1}Department of Physics, Huazhong Normal University, Wuhan 430079, China\\
\authormark{2}Center for Quantum Science and Technology, Jiangxi Normal University, Nanchang 330022, China\\
\authormark{3}School of Electronics and Information, Hangzhou Dianzi University, Hangzhou 310018, China\\}

\email{\authormark{*}gaox@mail.ccnu.edu.cn} 



\begin{abstract}	
The ultrastrongly coupling (USC) system has very important research significance in quantum simulation and quantum computing.  In this paper, the ultranarrow  spectrum  of a circuit QED system with two qubits ultrastrongly coupled  to a single-mode cavity  is studied. In the regime of USC, the JC model breaks down and the counter-rotating terms in the quantum Rabi Hamiltonian leads to the level anti-crossing in the energy spectrum. Choosing a single-photon driving field at the  point of avoided-level crossing, we can get an equivalent four-level  dressed state model, in which the dissipation of the two intermediate states is only related to the qubits decay. Due to the electron shelving of these two metastable states, a narrow peak appears in the  cavity emission spectrum. Furthermore, we find that the physical origin for the spectral narrowing is that the vacuum-induced quantum interference between  two transition  pathways couples the slow decaying incoherent components of the density matrix into the equations of the sidebands. This result  provides a possibility for the study of quantum interference effect in the USC system. 
\end{abstract}

\section{Introduction}\label{sec1}
Circuit quantum electrodynamics (QED) system provides an excellent experimental platform for the study of  quantum state engineering \cite{RevModPhys.73.357,Vion886,doi:10.1063/1.5089550}, quantum information processing \cite{Hennessy2007,Wendin_2017}, and quantum computing \cite{Chiorescu1869,Fedorov2012,Reed2012,Ofek2016}. Compared with cavity QED, the superconducting qubits in the circuit QED system are more strongly coupled with the cavity on the chip, leading to more frequent photon exchange and much shorter acquisition times in the experiment \cite{Haroche2020}. In 2010, the first  two experiments of ultrastrong coupling (USC) were realized in the circuit QED system \cite{Niemczyk2010,PhysRevLett.105.237001}. Since then, many new physical processes have  emerged in  the ultrastrong coupling circuit QED system, as degeneracy of vacuum \cite{PhysRevLett.104.023601}, modification of photon blockade \cite{PhysRevLett.109.193602},  nonclassical radiation from the thermal cavities \cite{PhysRevLett.110.163601}, and   vacuum-induced symmetry breaking \cite{PhysRevA.90.043817}.  In the USC regime,  the  existence of counter-rotating wave terms  makes the multi-photon Rabi oscillation possible, where multiple photons excite one qubit \cite{PhysRevA.92.063830}. Similarly, a single photon can simultaneously  excite multiple qubits \cite{PhysRevLett.117.043601}, which was soon verified by experiments\cite{PhysRevA.95.063848}.

Recently, analogous processes have been implemented in different USC systems to yield frequency conversion \cite{Kockum2017},  entanglement between photons \cite{PhysRevA.98.062327}, and spin squeezing \cite{PhysRevA.101.053818}. In the scenarios above, the exploration of the avoided-crossing region is particularly important.  At the avoided-crossing point, the single photon state and the multiple qubits state will be completely hybridized, resulting in a symmetric and anti-symmetric coupling state that does not exist under RWA. Surprisingly, different from the case of cavity-qubit near resonance \cite{oc283766,PhysRevA.91.043831}, when the frequency of the cavity mode is twice that of the qubit \cite{PhysRevLett.117.043601}, there will be two intermediate states containing only qubits and no photons. If we set the  cavity loss to be much greater than the qubits decay, we can get two metastable states with slow decay rate. Can these two metastable states bring us any interesting phenomena? We think that we can use the electron shelving   \cite{PhysRevA.52.3333,oc117560} of the intermediate states to realize the spectral narrowing.

Spectral narrowing can be applied to many research fields, such as laser spectroscopy \cite{Kues2017} and quantum sensing \cite{Khivrich2020}. As early as 1990, the sub-natural linewidth in the spontaneous emission spectrum of a three-level atom driven by two beams of light has been theoretically predicted by Narducci {\it et al.} \cite{PhysRevA.42.1630}, and a large number of experiments subsequently observed this  phenomenon \cite{PhysRevLett.66.2460,Vus,Changs,PhysRevLett.118.063601}.  
Recently, the ultranarrow linewidths have been observed in strongly coupled superconducting qubit systems \cite{PhysRevX.5.021035,PhysRevX.6.031004}.  In this paper, we investigate the ultranarrow  spectrum  of the qubit-cavity ultrastrong coupling system, which caused by the combined effect of quantum interference and  electron shelving. The system contains two metastable states, both of which correspond to narrow peaks in the emission spectrum, thus the narrowing effect in this paper is stronger than the previous work \cite{PhysRevA.93.033801}.

In this paper, we study the ultranarrow spectra and intensity-amplitude correlation of double superconducting  qubit-cavity coupling system in the regime of USC. Due to the counter-rotating wave terms in the quantum Rabi model (QRM), a splitting anti-crossing between  levels three and four appears in the  eigen-energy spectrum, as shown in Fig.~\ref{fig2}. Therefore, we choose the frequency of the driving field to   resonate with the transition from the ground state to the third energy level. And the circuit QED system is reduced to a four-level dressed state system.
At the point of  avoided-level crossing, an ultranarrow peak appears in the center of the cavity emission spectrum. Through careful analytical analysis, we find that the origin of ultranarrow linewidth is the slow decay of the incoherent terms in the density matrix and the quantum interference between two transition pathways. Moreover, at the point of level crossing, the extra inner sidebands of the emission spectrum will be highlighted compared to the  Mollow-like triplet.  Our results indicate that the spectral narrowing can be  achieved by  adjusting the coupling strength of the qubit-cavity ultrastrong coupling system.

This paper is organized as follows. In Section \ref{sec2}, we present the   model and  Hamiltonian of the circuit QED system in the USC regime, and then  show the level crossing and avoided-crossing in the eigen-energy spectrum. 	The  ultranarrow spectral line in the cavity emission spectrum is discussed in Section \ref{sec3}, including the five broad peaks and the ultranarrow peak imposed on the central peak. The main results and conclusions are summarized in Section \ref{sec5}.

\section{Model and system}~\label{sec2}	
We consider two identical superconducting qubits with transition frequency $\omega_{q}$ placed in a cavity on chip with resonance frequency $\omega_{c}$, and coupled together with coupling strength $g$. The coupling strength between each qubit and the single-mode cavity  is comparable to the cavity-qubit detuning $\Delta=\omega_{c}-\omega_{q}$, indicating the appearance of USC. In addition, the cavity is  driven by a coherent laser field with frequency $\omega_{l}$ and driving strength $\varepsilon$. The Hamiltonian describing the circuit QED system can be written as~\cite{PhysRevA.90.043817}
\begin{align}
H=H_{0}+H_{d},\label{c1}
\end{align}
where 
\begin{align}
H_{0}=\omega_{c}a^{\dag}a+\sum_{i}\left[ \frac{\omega_{q}}{2}\sigma_{z}^{(i)}+ g X \left(\cos\theta\sigma_{x}^{(i)}+\sin\theta\sigma_{z}^{(i)}\right) \right],\label{c2}
\end{align}
and the driving Hamiltonian can be expressed as~\cite{PhysRevA.91.043831}
\begin{align}
H_{d}=\varepsilon \cos(\omega_{l}t) X,\label{c3}
\end{align}
here $X=a+a^{\dag}$ is the cavity electric-field operator, $a$ and $a^{\dag}$ are the annihilation and creation operators for cavity photons. $\sigma_{x}^{(i)}$ and $\sigma_{z}^{(i)}$ are Pauli operators for the $i$th qubit.

In the regime of USC, the RWA breaks down and the Rabi Hamiltonian contains four counter-rotating terms of the form $\sigma_{+}^{(i)}a^{\dag}$,  $\sigma_{-}^{(i)}a$,  $\sigma_{z}^{(i)}a$, and $\sigma_{z}^{(i)}a^{\dag}$.
In general, it is a challenge to  obtain exact analytical solutions of energy levels for the USC systems.  And several approximation methods have been developed to deal with this kind of Hamiltonian. Such as third-order  perturbation theory \cite{PhysRevLett.117.043601} and adiabatic elimination method  \cite{PhysRevA.92.023842}. In order to get a more accurate expression of the effective Hamiltonian, we use the  method in Ref. \cite{PhysRevA.89.033827}.  By  diagonalizing $H_{0}$,  the numerical eigenvalues $E_{n}$ and eigenstates $\vert \psi_{n} \rangle $ which satisfy the stationary Schr\"odinger equation $H_{0}\vert \psi_{n} \rangle=E_{n}\vert \psi_{n} \rangle$ can be obtained. Then the  Rabi Hamiltonian  in Eq.~(\ref{c2}) can be replaced by
\begin{align}
H_{0}=\sum_{n=0}^\infty E_n\vert \psi_{n} \rangle\langle\psi_{n}\vert.\label{c4}
\end{align}

\begin{figure}[hbt]
\centering
\subfigure
{\label{fig2a}
\centering
\includegraphics[width=0.7\columnwidth]{fig2a}}
\subfigure
{\label{fig2b}
\centering
\includegraphics[width=0.35\columnwidth]{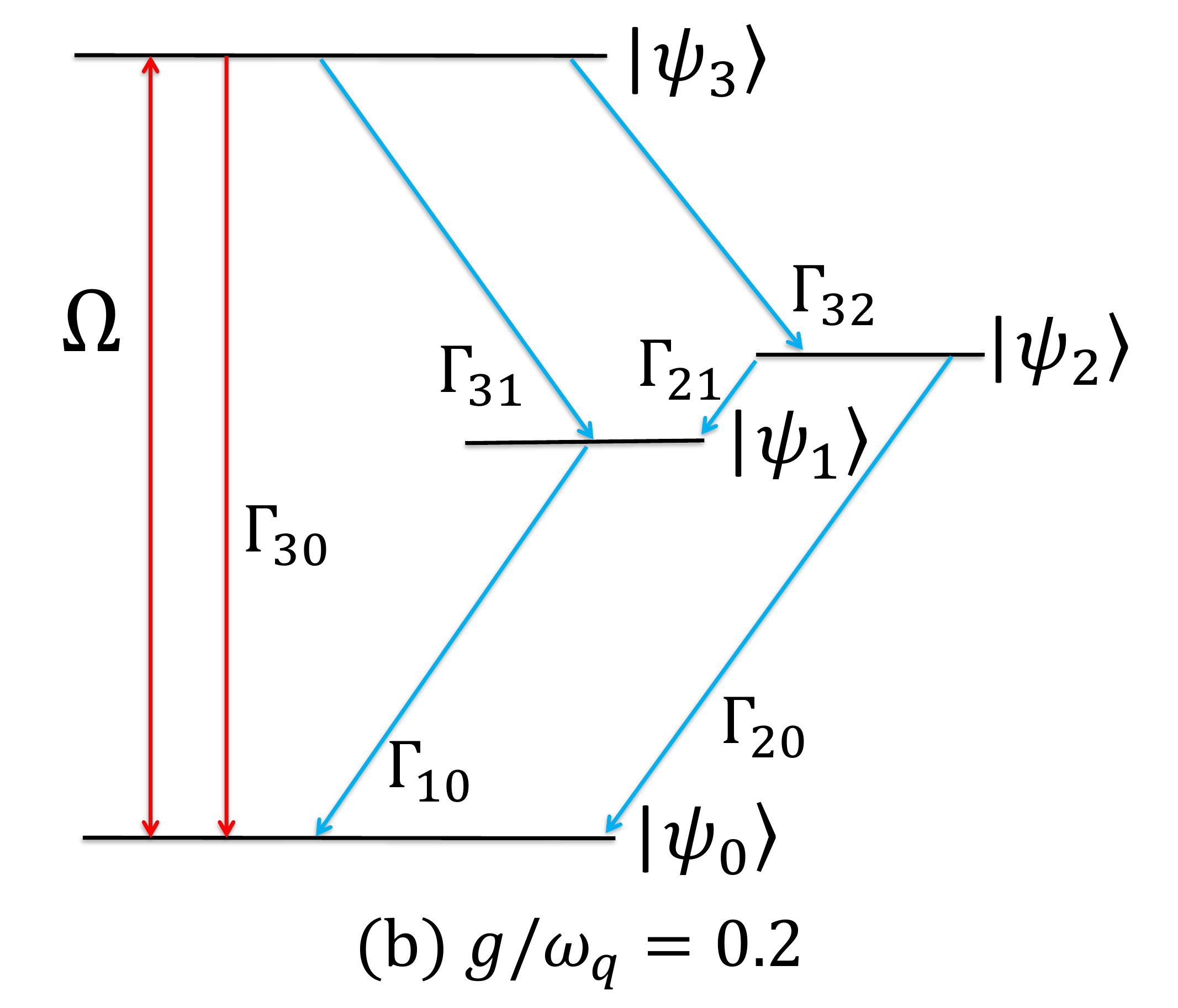}}
\subfigure
{\label{fig2c}
\centering
\includegraphics[width=0.338\columnwidth]{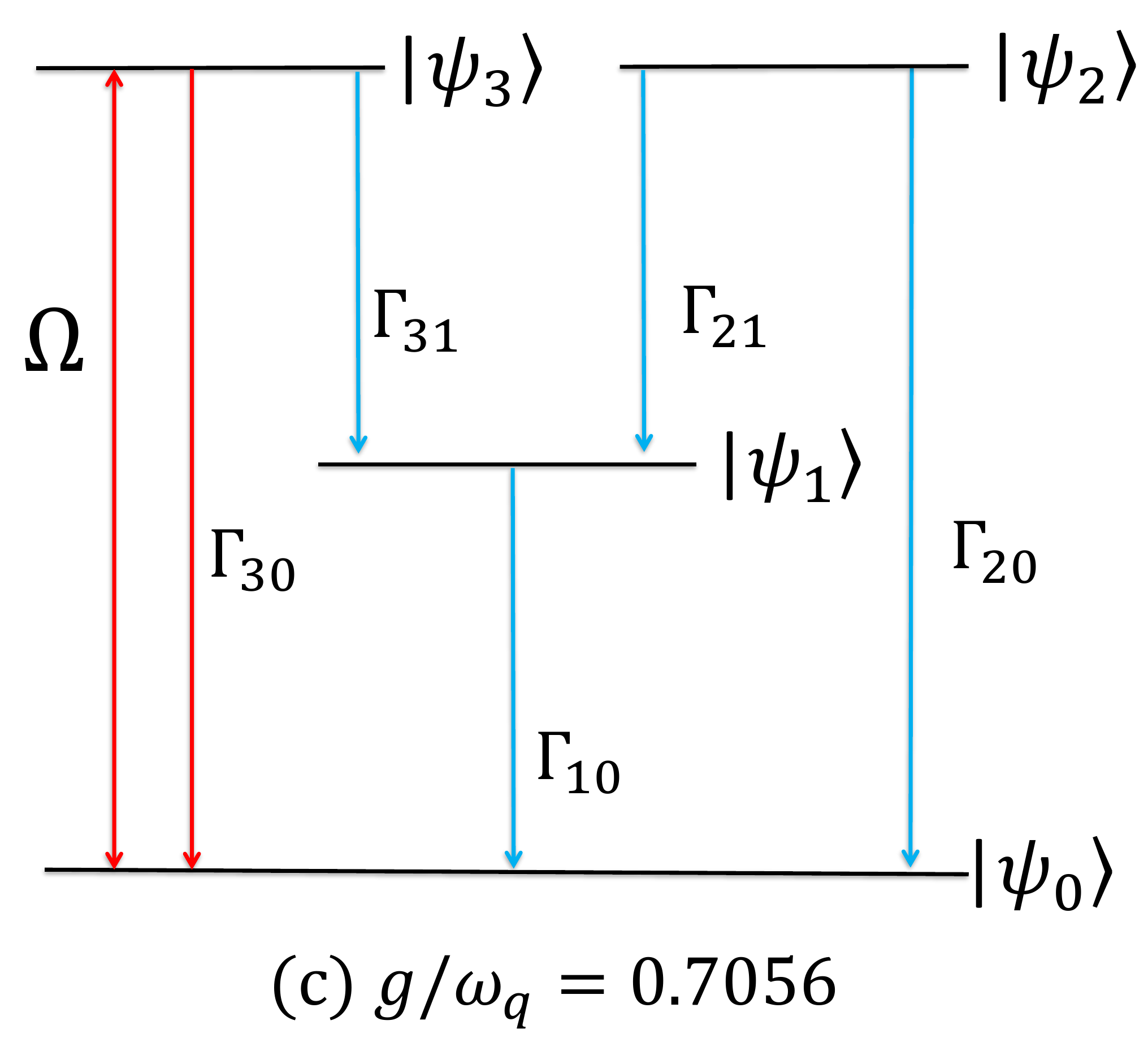}}
\caption{(a) Normalized energy ladders of the Rabi Hamiltonian $H_{0}$ as a function of the coupling strength $g$ for $\omega_{c}/\omega_{q}=1.915$ and $\theta=\pi/6$. (b), (c) The energy level scheme of  the cavity-qubits coupling system in  dressed state representation for  $g/\omega_{q}=0.2$ and $0.7056$, respectively. The modified Rabi frequency $\Omega$ stands for the standard amplitude of  driving field from ground state $\vert \psi_{0} \rangle$ to excited state $\vert \psi_{3} \rangle$, $\Gamma_{mn}\ (m>n)$ are the relaxation coefficients between each state. For the standard damping rates $\kappa\gg\gamma$, the dissipation is dominated by  $\Gamma_{30}$, which is marked in red.}
\label{fig2}
\end{figure}

Fig.~\ref{fig2a} illustrates the energy ladders of the Rabi Hamiltonian $H_{0}$ as a function of the cavity-qubit coupling strength $g$, where $\omega_{c}/\omega_{q}= 1.915$ and $\theta=\pi/6$. From Fig.~\ref{fig2a}, we can see the avoided-level crossing between $\vert\psi_{3}\rangle$ and $\vert\psi_{4}\rangle$ in the region around $g/\omega_{q}= 0.2$, which does not exist in the RWA. These two eigenstates  are approximate to  $( \vert e, e, 0\rangle  \pm \vert g, g, 1\rangle )/\sqrt{2}$. Obviously, the states  $\vert e, e, 0\rangle$ and
$\vert g, g, 1\rangle$ can only be coupled through the counter-rotating terms. In addition, energy levels two and three  crossing around $g/\omega_{q}=0.7$. These two regions [see arrows in Fig.~\ref{fig2a}] are  two cases that will be discussed in  this paper.

The driving Hamiltonian $H_{d}$  can be expanded by the dressed states $\vert \psi_{n} \rangle$, thus we have
\begin{align}
H_{d}=\varepsilon \cos(\omega_{l}t)\left[ \sum_{m,n>m}Z_{mn}\sigma_{mn} + H.c.\right],\label{c5}
\end{align}
where $Z_{mn}=\langle\psi_{m}\vert(a +a^{\dag})\vert \psi_{n} \rangle$ and $\sigma_{mn}=\vert \psi_{m} \rangle\langle\psi_{n}\vert$. Through the unitary transformation $H_{s}=e^{i H_{o}t} H_{d} e^{-i H_{o}t}$, we can obtain the system Hamiltonian
\begin{align}
H_{s}=\dfrac{\varepsilon}{2} \ \left[ \sum_{m,n>m} Z_{mn}\sigma_{mn}e^{-i(E_{nm}-\omega_{l})t}+ H.c. \right]  ,\label{c6}
\end{align}
where $E_{nm}=E_{n}-E_{m}$.

In order to further investigate the phenomenon of avoided-crossing, we therefore need to set up the driving from the ground state $\vert \psi_{0} \rangle$ to the target state $\vert \psi_{3} \rangle$. Hence, we choose the  frequency of diving field that resonates with the transition pathway $\vert \psi_{0} \rangle \rightarrow \vert \psi_{3} \rangle$ as
\begin{align}
\omega_{l}=E_{30}.\label{c7}
\end{align}
For the case of  $g/\omega_{q}=0.2$, the  detuning $\delta=E_{nm}-\omega_{l}$ in Eq.~(\ref{c6}) is minimized when $n=4$ and $m=0$, which corresponds to the splitting between levels three and four $E_{43}=2\times10^{-2}\omega_{q}$. However, the driving strength $\varepsilon$ we chose is weak, at the order of $10^{-3}\omega_{q}$, which is much small than $E_{43}$. According to the RWA, we can therefore ignore the fast oscillating terms in Eq.~(\ref{c6}) including $\sigma_{04}$. When $g/\omega_{q}=0.7056$, the dressed state $\vert \psi_{2} \rangle $ is also  driven resonantly by the external field. However, owing to the special form that $\vert\psi_{2}\rangle=( \vert g, e, 0\rangle  - \vert e, g, 0\rangle )/\sqrt{2}$, the element $Z_{02}$, which corresponds to the transition  from $\vert \psi_{0} \rangle$ to $\vert \psi_{2} \rangle $ is equal to  zero. This  causes the vanish of $\sigma_{02}$ in Eq.~(\ref{c6}). Ultimately, in the limit of weak driving, the effective  Hamiltonian   can be simplified to
\begin{align}
H_{s}=\dfrac{\Omega}{2}\ (\sigma_{03}+\sigma_{30}),\label{c9}
\end{align}
where $\Omega=\varepsilon \langle\psi_{0}\vert ( a + a^{\dag}) \vert \psi_{3} \rangle$.

Therefore, we can make a four-state truncation of the Hilbert space. And the dressed state structures of the two cases  we care about are shown in Fig.~\ref{fig2b} and \ref{fig2c}. Here we only consider that the system interacting with zero-temperature baths. Owing to the effects of the counter-rotating terms, the system can be driven out of the ground state, thus the approach of standard quantum-optical master equation breaks down.  Assuming a weak coupling of the system and the baths, the dissipations can be treated by the Born-Markov approximation. Thus the modified master equation for the reduced density matrix can be obtained as \cite{PhysRevLett.110.163601,PhysRevA.88.063829}
\begin{align}
\dot{\rho}(t)=-i[H_{s},\rho(t)]+{\cal L}_{c}\rho(t)+\sum_{i=1,2}{\cal L}_{a}^{(i)}\rho(t),\label{c10}
\end{align}
where ${\cal L}_{a}^{(i)}$ and ${\cal L}_{c}$ are Liouvillian superoperators which describing the dissipations of the qubits and the cavity mode respectively. And ${\cal L}_{x}\rho(t)=\sum_{j,k>j}\Gamma^{jk}_{x}{\cal D}[\vert \psi_{j }\rangle\langle \psi_{k }\vert]\rho(t)$, for $x=c,\sigma^{(i)}_{-}$. The superoperator ${\cal D}$  is defined as ${\cal D}[O]\rho=\frac{1}{2}(2\,O\,\rho\,O^{\dag}-\rho\,O^{\dag}\,O-O^{\dag}\,O\,\rho)$. The relaxation coefficients of the qubits and the cavity mode are defined as
\begin{equation}
\begin{aligned}
\Gamma^{jk}_{x}&\equiv\gamma_{i}\dfrac{E_{jk}}{\omega_{q}}\vert\langle \psi_{k }\vert( \sigma_{-}^{(j)} - \sigma_{+}^{(i)})\vert \psi_{j }\rangle\vert^{2},~~(x=\sigma^{(i)}_{-}),\\
\Gamma^{jk}_{x}&\equiv\kappa \ \dfrac{E_{jk}}{\omega_{c}}\vert\langle \psi_{k }\vert(\ a \ \  - \ \  a^{\dag}\ )\vert \psi_{j }\rangle\vert^{2},~~(x=c),\label{c11}
\end{aligned}
\end{equation}
where $\gamma_{i}$  and $\kappa$ are standard damping rates of the qubits and cavity, and here we set $\gamma_{1}=\gamma_{2}=\gamma$. 

\begin{figure}[htb]
\centering
\includegraphics[width=0.75\columnwidth]{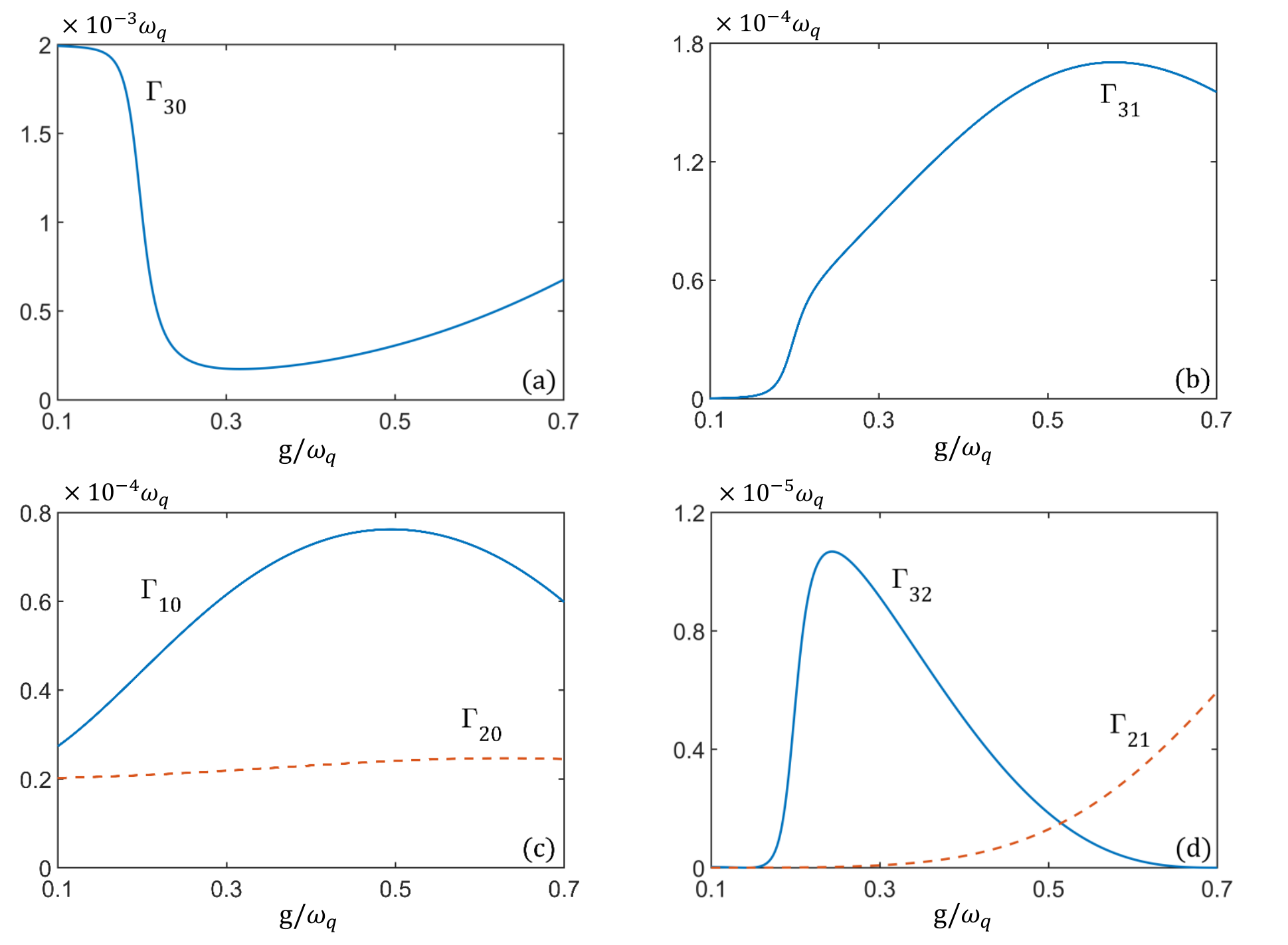}
\caption{(a)-(d) The relaxation coefficients $\Gamma_{mn}$ as functions of the coupling strengths $g$ for $\omega_{c}/\omega_{q}=1.915$ and $\theta=\pi/6$.   The standard damping rates of the qubits and cavity are $\kappa=2\times10^{-3}\omega_{q}$, $\gamma=2\times10^{-5}\omega_{q}$.}
\label{fig3}
\end{figure}

The relaxation coefficients between each two dressed state for different coupling strengths are shown in Fig.~\ref{fig3}. When the coupling between  qubits and cavity is weak, the interaction Hamiltonian in Eq.~(\ref{c2}) can be treated as perturbation, and the RWA is available. Consequently, the relaxation coefficients are reduced to the standard damping rates of the qubits and cavity in the case of weak coupling, where $\Gamma_{30}=\kappa$, $\Gamma_{10}=\Gamma_{20}=\gamma$. With the increase of $g$, the RWA breaks down and the counter-rotating terms are taken into account, which results in variations of the relaxation coefficients. In addition, at the point of avoided-level crossing that $g/\omega_{q}=0.2$, the eigenstates $\vert\psi_1\rangle$ and $\vert\psi_2\rangle$ are approximate to  $( \vert g, e, 0\rangle  \pm \vert e, g, 0\rangle )/\sqrt{2}$. For the case of $\kappa\gg\gamma$, $\vert\psi_1\rangle$ and $\vert\psi_2\rangle$ contain only qubits and no photons. These two metastable states are very important to us, and it is the electron shelving on these two states that causes the spectral narrowing. It needs to be emphasized that, different from the long-lived metastable states in Ref. \cite{PhysRevA.95.023829}, the cascaded transition in this paper differs by several orders of magnitude from the cascaded transition, resulting in a longer lifetime of metastable state here.

\section{Ultranarrow spectral line in the cavity emission spectrum}\label{sec3}
According to the input-output relations  proposed by Ridolfo {\it et al.}\cite{PhysRevLett.109.193602,PhysRevLett.110.163601},
the incoherent cavity emission spectrum $S(\omega)$ in the USC regime  is defined as
\begin{align}
S_{inc}(\omega)\propto \lim\limits_{t\rightarrow\infty}\ 2{\cal R}\int_{0}^{\infty}\langle\delta\dot{X}^{-}(t)\delta\dot{X}^{+}(t+\tau)\rangle e^{i \omega \tau }d\tau,\label{c14}
\end{align}
where $\dot{X}=-i X_{0}(a-a^{\dag})$ and ${\cal R}$ denotes the real part. Note that the   annihilation (creation) operator $a$ ($a^{\dag}$) in the standard input-output relations under RWA are replaced by $\dot{X}^{+}$ ($\dot{X}^{-}$), which represent the positive (negative) frequency components of $\dot{X}$. And $X_{0}$ is the rms zero-point field amplitude which is assumed to be unit in this paper.
By expanding $\dot{X}$ in the dressed state basis $\vert \psi_{i}\rangle$,  $\dot{X}^{+}$ can be expressed as $\dot{X}^{+}=-i\sum_{m,n>m}E_{nm}X_{mn}\sigma_{mn}$, where $X_{mn}=\langle\psi_{m}\vert \dot{X} \vert \psi_{n}\rangle$ and $\dot{X}^{-}=(\dot{X}^{+})^{\dag}$. For the four-level dressed states system in Fig.~\ref{fig2b} and \ref{fig2c},   we can derive the expression as
\begin{align}
\dot{X}^{+}(t)=\alpha_{01}\sigma_{01}(t)+\alpha_{03}\sigma_{03}(t)+\alpha_{13}\sigma_{13}(t),\label{c15}
\end{align}
where $\alpha_{mn}=-E_{nm}\langle\psi_{m}\vert( a-a^{\dag} )\vert \psi_{n}\rangle$.

Since $\rho_{01}^{ss}=\rho_{13}^{ss}=0$ (see Supplement 1, S1), the initial values of the cross-correlations in Eq~(\ref{c14}) are equal to zero, which leads to the disappearance of the cross-correlations. Therefore, the  emission spectrum can be simplified as three auto-correlation functions of the transition operators in Eq~(\ref{c15})
\begin{align}
S_{inc}(\omega)=S_{1}(\omega)+S_{2}(\omega)+S_{3}(\omega),\label{c16}
\end{align}
where
\begin{align}
S_{1}(\omega)=&\vert\alpha_{03}\vert^2\lim\limits_{t\rightarrow\infty}\ 2{\cal R}\int_{0}^{\infty}\langle\delta\sigma_{30}(t)\delta\sigma_{03}(t+\tau)\rangle e^{i \omega \tau }d\tau,\nonumber\\
S_{2}(\omega)=&\vert\alpha_{01}\vert^2\lim\limits_{t\rightarrow\infty}\ 2{\cal R}\int_{0}^{\infty}
\langle\delta\sigma_{10}(t)\delta\sigma_{01}(t+\tau)\rangle e^{i \omega \tau }d\tau\nonumber\\
S_{3}(\omega)=&\vert\alpha_{13}\vert^2\lim\limits_{t\rightarrow\infty}\ 2{\cal R}\int_{0}^{\infty}\langle\delta\sigma_{31}(t)\delta\sigma_{13}(t+\tau)\rangle e^{i \omega \tau }d\tau.\label{c17}
\end{align}

\begin{figure}[htb]
\centering
\includegraphics[width=0.75\columnwidth]{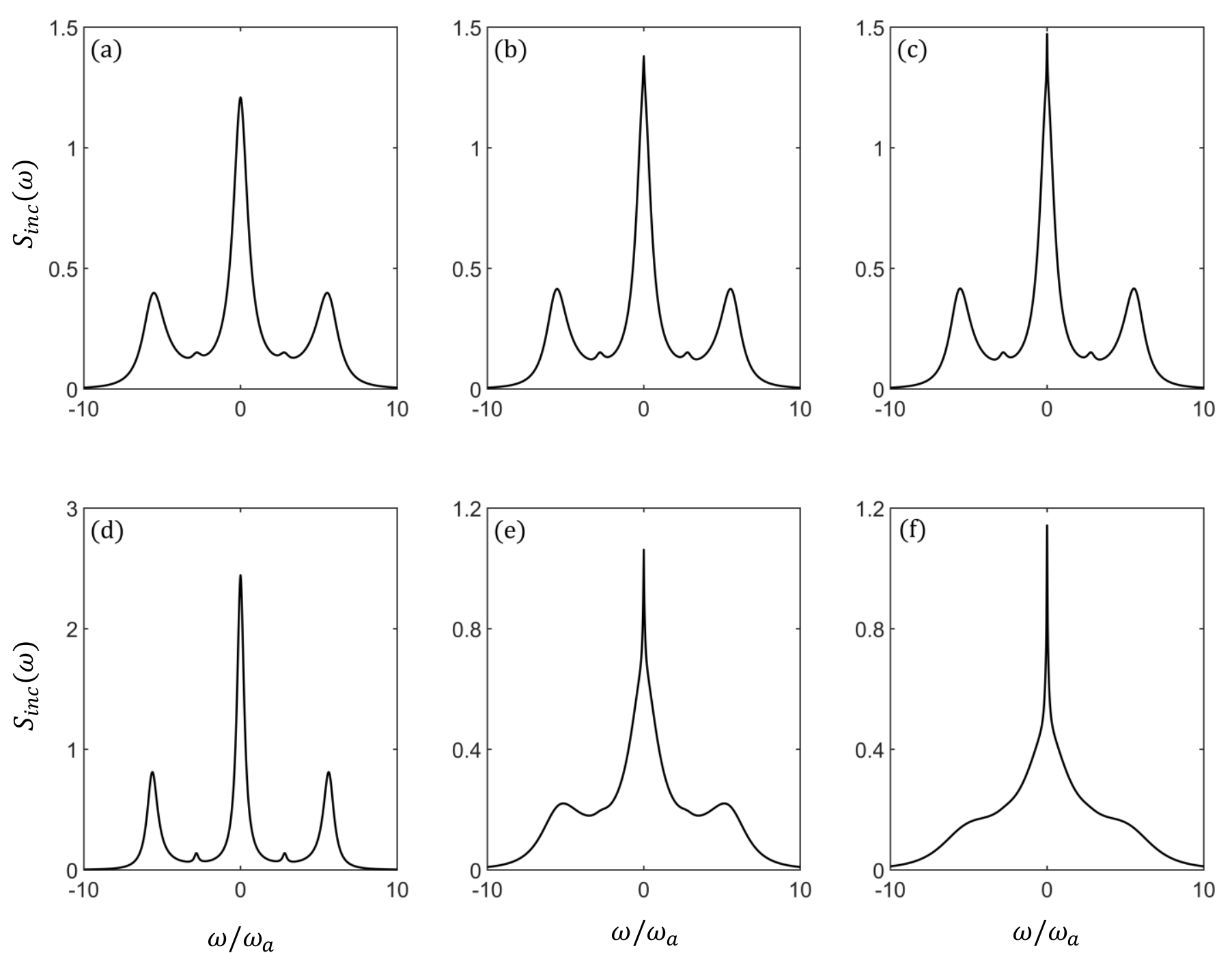}
\caption{The incoherent emission spectrum $S_{inc}(\omega)$ of the reduced nondegenerate four-level system in Fig.~\ref{fig2b}, with $g/\omega_{q}=0.2$, $\omega_{c}/\omega_{q}=1.915$, $\theta=\pi/6$, and $\varepsilon/\omega_{a}=8$, for (a) $\kappa=2$, $\gamma=0.1$; (b) $\kappa=2$, $\gamma=0.02$; (c) $\kappa=2$, $\gamma=0.01$; (d) $\kappa=1$, $\gamma=0.02$; (e) $\kappa=4$, $\gamma=0.02$; and (f) $\kappa=6$, $\gamma=0.02$ (units of $\omega_{a}$ where $\omega_{a}= 10^{-3}\omega_{q}$).}
\label{fig4}
\end{figure}

 We first consider the case of avoided-level crossing that $g/\omega_{q}=0.2$, the spectra of the  USC system  are shown in Fig.~\ref{fig4}. One finds that the incoherent  emission spectrum consists of a Mollow-like triplet \cite{mollow}, two additional symmetrically sidebands, and most importantly, a ultranarrow peak imposed on the central peak. Moreover, it is clear that the height and width of the Mollow-like triplet  hardly changed with the decay rate of the qubits $\gamma$, as shown in Figs.~\ref{fig4}(a)-(c). However, as the cavity dissipation $\kappa$ increases, the linewidths of the Mollow-like triplet grows and the height reduces, as shown in Figs.~\ref{fig4}(d)-(f).  Meanwhile, the relative linewidth of the narrow peak out of the central peak is proportional to $\gamma$ and inversely proportional to $\kappa$.

In order to better explain the physical origin of spectral narrowing in the emission spectrum, we diagonalize the effective  Hamiltonian in Eq~(\ref{c9}) to get the dressed states
\begin{align}
\begin{aligned}
\vert + \rangle=\frac{1}{\sqrt{2}}(\vert \psi_{3} \rangle + \vert \psi_{0} \rangle),\\
\vert - \rangle=\frac{1}{\sqrt{2}}(\vert \psi_{3} \rangle - \vert \psi_{0} \rangle),\label{c18}
\end{aligned}
\end{align}
thus the effective  Hamiltonian can be expressed as
\begin{align}
\begin{aligned}
H_{s}=\dfrac{\Omega}{2}\ (\sigma_{++}-\sigma_{--}).\label{c19}
\end{aligned}
\end{align}

The two-time correlation $lim_{t\rightarrow\infty}\langle\delta\sigma_{mn}(t)\delta\sigma_{nm}(t+\tau)\rangle$ can be obtained by invoking the quantum regression theorem with the equations of motion in the dressed state representation.  The detailed derivation of the analytical results for emission spectrum is given in Supplement 1, S2. It should be emphasized that secular approximation cannot be used in the derivation of ultranarrow peak. Here we present two equations under secular approximation and non-secular approximation, which correspond to the central peak and narrow peak respectively
\begin{subequations}
\begin{align}
\dfrac{d}{dt}\left(\rho_{++}-\rho_{--}\right)=&-\dfrac{A}{2}\left( \rho_{++}-\rho_{--}\right),\label{c23a}\\
\dfrac{d}{dt}\left(\rho_{++}+\rho_{--}\right)=&-\left(\Gamma_{12}^++\Gamma_{23}^+\right)\left( \rho_{++}+\rho_{--}\right) -\Gamma_{23}^+\left(\rho_{-+}+\rho_{+-}\right)+\Gamma_{12}^-\left(\rho_{11}-\rho_{22}\right),\label{c23b}
\end{align}\label{c23}
\end{subequations}
where $\Gamma_{12}^\pm=(\Gamma_{10}\pm\Gamma_{20})/2$ and $\Gamma_{23}^\pm=(\Gamma_{31}\pm\Gamma_{32})/2$. And eventually we can obtain  the analytical expression of the spectrum as 
\begin{subequations}
\begin{align}
S_1(\omega)=&\dfrac{\vert\alpha_{03}\vert^2}{2}{\cal R}\left. \bigg[ \right.\dfrac{C_0}{\lambda_0-i\omega}+\dfrac{C_0^+}{\lambda_0^+-i\omega}+\dfrac{C_0^-}{\lambda_0^--i\omega}+\dfrac{C_1^+}{\lambda_1^+-i\omega}+\dfrac{C_1^-}{\lambda_1^--i\omega}\left.\bigg]\right.,\label{c20a}\\
S_2(\omega)=&\vert\alpha_{01}\vert^2{\cal R}\left[ \dfrac{\rho_{11}^{ss}}{\lambda_2^+-i\omega}+\dfrac{\rho_{11}^{ss}}{\lambda_2^--i\omega}\right],\label{c20b}\\
S_3(\omega)=&\vert\alpha_{13}\vert^2{\cal R}\left[ \dfrac{\rho_{++}^{ss}}{\lambda_2^+-i\omega}+\dfrac{\rho_{--}^{ss}}{\lambda_2^--i\omega}\right],\label{c20c}
\end{align}\label{c20}
\end{subequations}
where the eigenvalues of the  coefficient matrix for the master equation are
\begin{subequations}
\begin{align}
\lambda_0=&\dfrac{A}{2},\label{c21a}\\
\lambda_0^\pm=&\dfrac{2A+\Gamma_{30}}{4}\pm i\Omega,\label{c21b}\\
\lambda_1^\pm=&\dfrac{D\pm\Delta}{4},\label{c21c}\\
\lambda_2^\pm=&\dfrac{A+2\Gamma_{10}}{4}\pm \dfrac{i\Omega}{2},\label{c21d}
\end{align}\label{c21}
\end{subequations}
and the amplitude factors are
\begin{align}
C_0=&\rho_{++}^{ss}+\rho_{--}^{ss},\qquad C_0^\pm=\rho_{\pm\pm}^{ss},\nonumber\\
C_1^\pm=&\dfrac{i \Gamma_{30}(\rho_{-+}^{ss}-\rho_{+-}^{ss})}{2 \Omega } \left. \bigg[ \right.\left( 1\pm\dfrac{\Gamma_{+}+\Gamma_{32}}{\Delta}\right)\rho_{11}^{ss}+\left(  1\pm\dfrac{\Gamma_{-}+\Gamma_{32}}{\Delta}\right) \rho_{22}^{ss}  \left.\bigg]\right.,\label{c22}
\end{align}
here we have $\Delta=\sqrt{\Gamma_{+}^2+2\Gamma_{-}\Gamma_{32}+\Gamma_{32}^2}$, and $\Gamma_{\pm}=\Gamma_{31}-\Gamma_{21}\pm 2(\Gamma_{10}-\Gamma_{20})$, $D=2\Gamma_{10}+2\Gamma_{20}+\Gamma_{21}+\Gamma_{31}+\Gamma_{32}$.

\begin{figure}[htb]
\centering
\includegraphics[width=0.7\columnwidth]{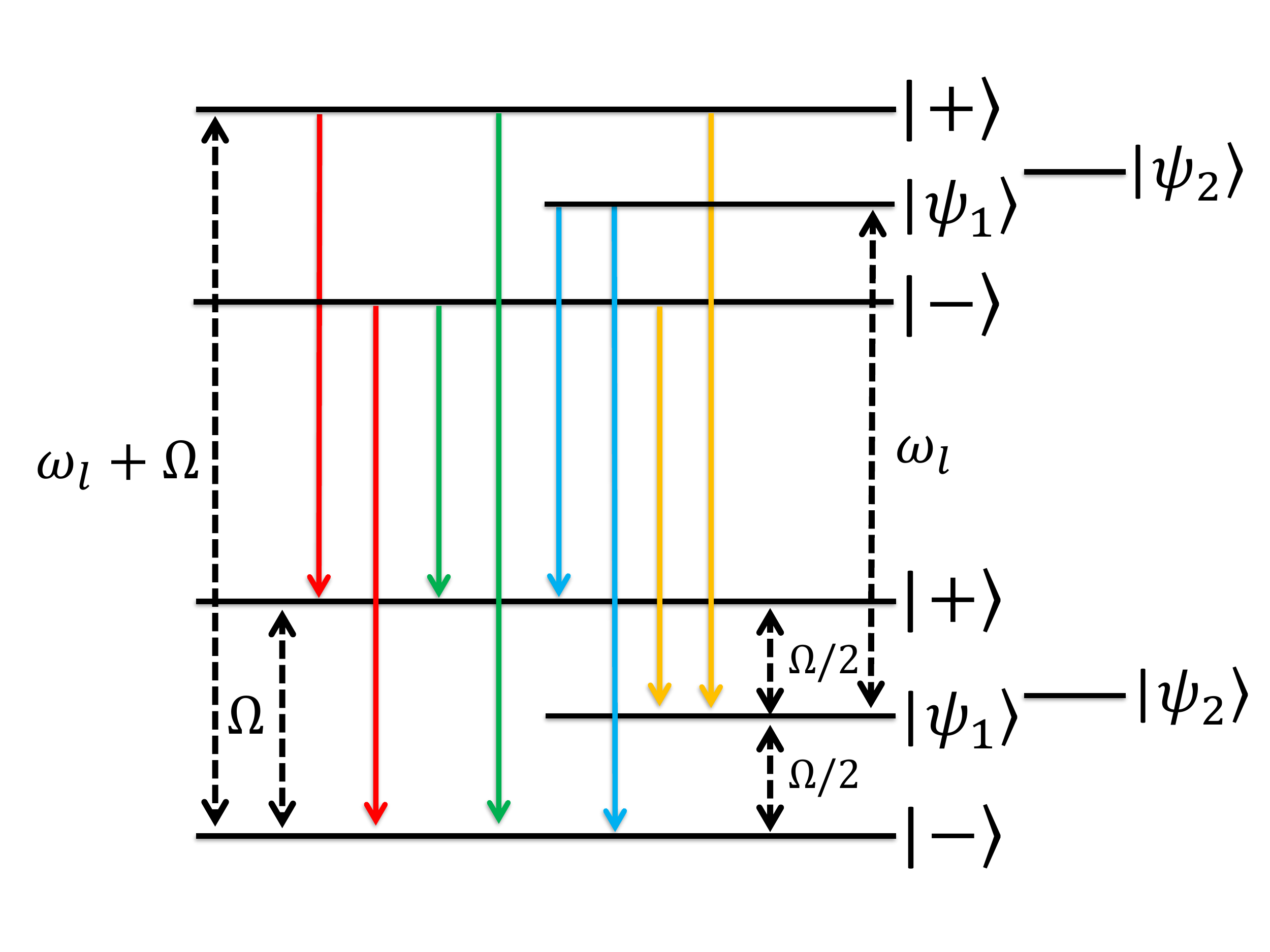}
\caption{Diagram of dressed states and dressed state transition pathways of each component in the  emission spectrum.}
\label{fig13}
\end{figure}

The  emission spectrum consists of six parts:  the central peak of the Mollow-like triple, the outer sidebands located at $\pm\Omega$, and the inner sidebands located at $\pm\Omega/2$. And most importantly, the additional narrow peak at line center which is significantly sharper than any other peaks. The physical origin of each component  can be understood through the transition channels between the dressed states from two contiguous manifolds,  as shown in Fig.~\ref{fig13}. 

The central peak originates from  transition channels $\vert+\rangle\rightarrow\vert+\rangle$ and $\vert-\rangle\rightarrow\vert-\rangle$. The corresponding Lorentzian with linewidth $A$  is related to the decay rate of the the population difference $\rho_{++}-\rho_{--}$. The outer sidebands located at $\pm\Omega$ come from transition pathways $\vert+\rangle\rightarrow\vert-\rangle$ and $\vert-\rangle\rightarrow\vert+\rangle$. The Lorentzians with linewidth $(2A+\Gamma_{30})/2$ are associated with the decay rates  of the dressed state coherences $\rho_{+-}$ and $\rho_{-+}$, respectively. Since the stationary solutions  $\rho_{++}^{ss}$ and $\rho_{--}^{ss}$ are equal,  the heights of the two outer sidebands are the same. Moreover,  the inner sidebands located at $\pm\Omega/2$ come from spectra $S_2(\omega)$ and $S_3(\omega)$.  In the spectrum of $S_2(\omega)$, the right (left) side peak located at $\Omega/2$ ($-\Omega/2$) is  resulting from the decay of $\vert\psi_{1}\rangle\rightarrow\vert-\rangle$ ($\vert+\rangle$),  with the weight $\rho_{11}$, as shown in the blue arrows of  Fig.~\ref{fig13}. However, in the spectrum of $S_3(\omega)$, the right side peak is   resulting from the transition channel $\vert+\rangle\rightarrow\vert\psi_{1}\rangle$, with the weight $\rho_{++}$. And the left side peak  is  resulting from the transition channel  $\vert-\rangle\rightarrow\vert\psi_{1}\rangle$, with the weight $\rho_{--}$, as shown in the yellow arrows of Fig.~\ref{fig13}.  Owing to the same steady state population of $\rho_{++}$ and $\rho_{--}$, the heights of the inner sidebands are the same. 

In summary, the transition channels corresponding to the five peaks mentioned above are either start from $\vert\pm\rangle$ or end to the $\vert\pm\rangle$, so their linewidths are related to $A$. Since $A$ contains  a dominant dissipation process with relaxation coefficient $\Gamma_{30}$, these five peaks are all broad peaks.

Particularly, the extra narrow peak is an additional structure observed when we detecting the outer sidebands of the cavity emission spectrum. Since the  stationary solution $\rho_{+-}^{ss}=\rho_{-+}^{ss}=0$ under secular approximation, i.e., $C_1^\pm=0$,  the narrow peak will disappear, thus we  cannot apply secular approximation approach in the process of solving the narrow peak. Through the analytical analysis of the ultranarrow peak in Supplement 1, S2B, we find that the narrow peak  located at the line center comes from the correlation between the central peak and the side peaks of Mollow-like triple, and the corresponding motion  equations are coupled to the equations of the sidebands through vacuum-induced quantum interference, as shown in Supplement 1, Eq.~(S17). When we detecting the fluorescence radiated by the transition pathways $\vert \pm \rangle\rightarrow\vert \mp \rangle$, which corresponds to the observation operator $\langle\delta\sigma_{+-}-\delta\sigma_{-+}\rangle$, we can see not only the sidebands, but also an ultranarrow peak imposed on the central peak.

Comparing Eq.~(\ref{c23b}) with Eq.~(\ref{c23a}), we find that the time evolution of the population difference $\rho_{++}-\rho_{--}$ corresponding to the central peak is a rapid decay process with decay rate $A/2$, while the incoherent injection $\rho_{++}+\rho_{--}$ corresponding to the narrow peak is a slow decay process. If there is no electron shelving of the intermediate states $\vert \psi_1 \rangle$ and $\vert \psi_2 \rangle$, such as two-level atomic system, then the incoherent injection will be a constant that does not evolve with time, and  there will be no narrow peak. Furthermore, the narrow peak here is different from the narrow peaks in the V-type  three-level atomic systems~\cite{PhysRevLett.77.3995,PhysRevA.56.3011}.  In  those systems, the narrow peaks at the line center are directly  derived from the motion equation of the observation operator $\langle\delta\sigma_{++}-\delta\sigma_{--}\rangle$. However, here we find the narrow peak imposed on the central peak when detecting the fluorescence spectrum of the sidebands, which corresponds to observation operator $\langle\delta\sigma_{+-}-\delta\sigma_{-+}\rangle$. In previous works \cite{PhysRevLett.77.3995,PhysRevA.56.3011}, in order to observe the quantum interference between different transition pathways through spectral narrowing, the two upper energy states must be degenerated or nearly degenerated. In this way, the frequencies corresponding to the transitions from the two upper states to the ground state  are at the same level, and the quantum interference effect between these two transition channels is preserved after the secular approximation. However, in this paper, we can achieve spectral narrowing in three-level ($g/\omega_{q}=0.7056$) or four-level ($g/\omega_{q}=0.2$) systems without degenerate states, and dig out the hidden vacuum-induced quantum interference effects via spectral  narrowing.  This finding  opens up  a new possibility for studying the quantum interference effect in the USC system.
\begin{figure}[htb]
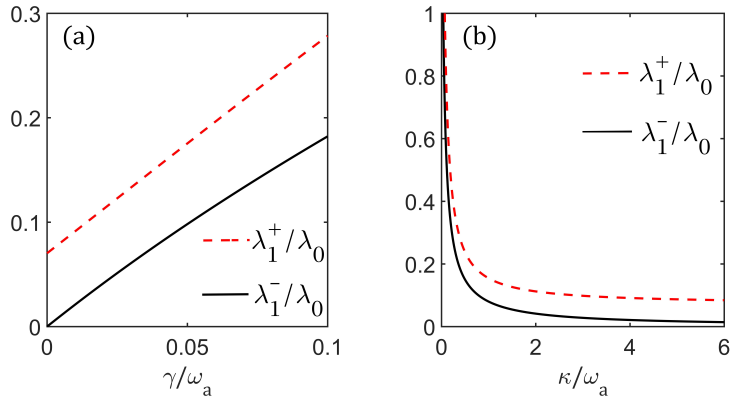

\centering
\subfigure
{\label{fig5a}
\centering
\includegraphics[width=0.36\columnwidth]{fig5a}}
\subfigure
{\label{fig5b}
\centering
\includegraphics[width=0.385\columnwidth]{fig5b}}
\caption{The relative linewidths of the narrow peak out of the central peak as functions of (a) $\gamma$ ($\kappa/\omega_a=2$); and (b) $\kappa$ ($\gamma/\omega_a=0.02$). The other parameters are the same as Fig.~\ref{fig4}.}
\label{fig5}
\end{figure}

Since $\kappa\gg \gamma$, it  is clear from Eq.~(\ref{c21}) that the linewidths in $\lambda_0$, $\lambda_0^\pm$ and $\lambda_2^\pm$ which containing factor $A$ increases with $\kappa$. Meanwhile, these linewidths does not change with $\gamma$.
But the narrow linewidths $\lambda_1^\pm$ are more sensitive to $\gamma$. As shown in Fig.~\ref{fig5}, the relative linewidth of the narrow peak out of the central peak is proportional to $\gamma$ and inversely proportional to $\kappa$. From this figure, we can see that the minimum value of $\lambda_1^+$ (dashed red curve) is about $0.1\lambda_0$, which is not small enough to observe the narrowing of the spectrum. Actually, the occurrence of ultra-narrow spectrum is mainly due to the ultra-small relative linewidth $\lambda_1^-/\lambda_0$ (solid black curve).  For $\kappa/\omega_a=2$ and $\gamma/\omega_a=0.1$, the linewidth $\lambda_1^-$ is $1/5$  of the  center peak linewidth  $\lambda_0$, while the height of the narrow peak is $1/16$ of the center peak,  thus the spectral narrowing is  not obvious in   Fig.~\ref{fig4}(a). However, when $\gamma\ll\kappa$, one can obtain a sharp ultranarrow peak with the  linewidth of $0.01\lambda_0$, meanwhile, the height of the narrow peak is $1/4$ of the center peak. From Eq.~(\ref{c22}), we can see that the amplitude $C_1^-$ of the narrow peak is multiplied by a factor $\Gamma_{30}/\Omega$ compared to the amplitude of the sidebands $C_0^\pm$. This factor is a small quantity, thus the narrow peak can be regarded as a correction to the outer sidebands. However, the linewidth $\lambda_1^-$ is also a  small quantity, which makes the height of the narrow peak  $C_1^-/\lambda_1^-$ comparable to the height of the central peak. At this point, the ultranarrow peak  is highlighted in the emission spectrum, as shown in Figs.~\ref{fig4}(c), \ref{fig4}(e) and \ref{fig4}(f).

\begin{figure}[htb]
\centering
\includegraphics[width=0.75\columnwidth]{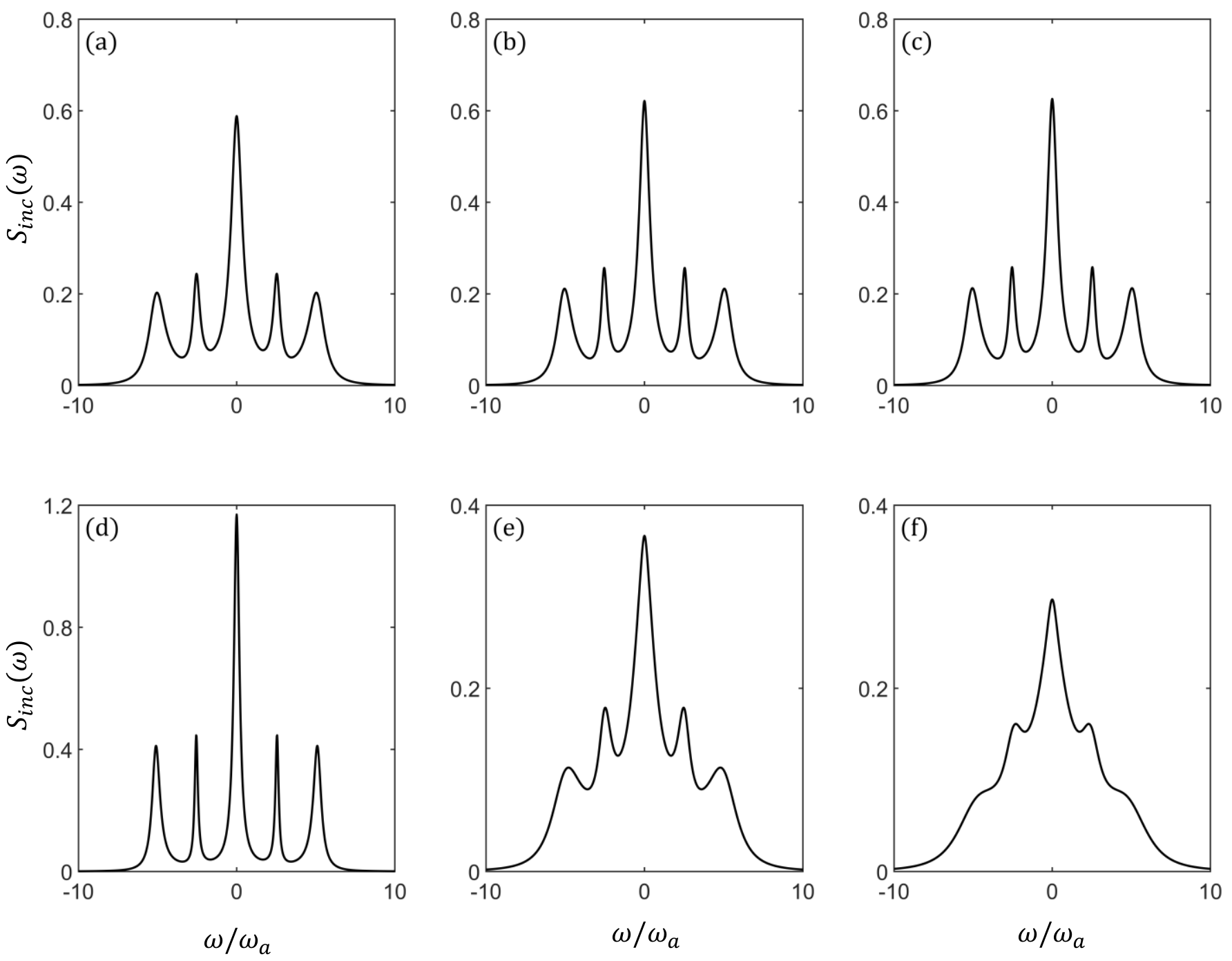}
\caption{The incoherent emission spectrum $S_{inc}(\omega)$ of the reduced degenerate four-level system as shown in Fig.~\ref{fig2c}. The other parameters are the same as Fig.~\ref{fig4}, except $g/\omega_q=0.7056$.}
\label{fig6}
\end{figure}

We next consider the case of level crossing that $g/\omega_q=0.7056$, the incoherent  emission spectra are shown in Fig.~\ref{fig6}. As before, the Mollow-like triplet and the additional sidebands hardly changed with $\gamma$, but the linewidths of them  grows and the height reduces with the increase of $\kappa$. However, different from the previous situation, the two inner sidebands are more pronounced, and the narrow peak is less obvious in this case.

\begin{figure}[htb]
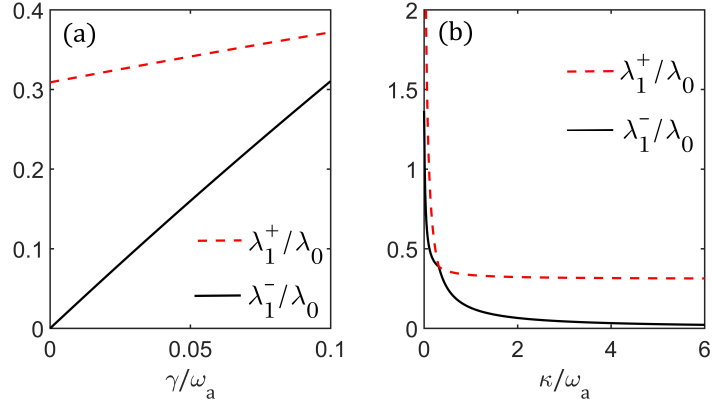

\centering
\subfigure
{\label{fig7a}
\centering
\includegraphics[width=0.36\columnwidth]{fig7a}}
\subfigure
{\label{fig7b}
\centering
\includegraphics[width=0.36\columnwidth]{fig7b}}
\caption{The relative linewidths of the narrow peak out of the central peak as functions of (a) $\gamma$ ($\kappa/\omega_a=2$); and (b) $\kappa$ ($\gamma/\omega_a=0.02$). The other parameters are the same as Fig.~\ref{fig4}, except $g/\omega_q=0.7056$.}
\label{fig7}
\end{figure}

By analyzing the components of dressed states $\vert\psi_1\rangle$ and $\vert\psi_3\rangle$, we find that when $g$ increases from $0.2\omega_{q}$ to $0.7056\omega_{q}$,  the probability amplitudes of the bare states with photon population (such as $\vert e,g,1\rangle$ and $\vert g,g,2 \rangle$) increase. This leads to increased photon exchange between $\vert\psi_1\rangle$ and $\vert\psi_3\rangle$, i.e., the coefficient of $\kappa$ in $\Gamma_{31}$ increases. Since $\kappa\gg\gamma$, $\Gamma_{31}$ increases and the population accumulates on $\rho_{11}$. In this way, the transition pathways  $\vert\psi_1\rangle\rightarrow\vert\pm\rangle$ and $\vert\pm\rangle\rightarrow\vert\psi_1\rangle$ that associated with $S_2(\omega)$ and $S_3(\omega)$ are enhanced, as shown in Fig.~\ref{fig13}. Thus the proportion of $S_2(\omega)$ and $S_3(\omega)$ in the entire spectrum increases, and finally the inner sidebands are highlighted. Meanwhile, the increase of $\Gamma_{31}$ means that the decay rate of $\rho_{++}+\rho_{--}$ in Eq.~(\ref{c23b})  increases, and the linewidth $\lambda_1^+$ corresponding to the narrow peak is getting bigger, as shown in Fig.~\ref{fig7a}. Whereas the ultra-small linewidth $\lambda_1^-/\lambda_0$ (solid black curve) corresponding to the ultra-narrow peak   is basically the same as before. So what makes the narrow peak to become inconspicuous? We can revisit the energy level scheme in Fig.~\ref{fig2}, the difference between Fig.~\ref{fig2b} and \ref{fig2c} is that when $g/\omega_q=0.7056$, the energy states $\vert\psi_2\rangle$ and $\vert\psi_3\rangle$  degeneracy, which makes the relaxation coefficient $\Gamma_{32}=0$. Under this condition, one obtain that $\rho_{22}^{ss}=0$, and
\begin{align}
\lambda_1^+=&\dfrac{2\Gamma_{10}+\Gamma_{31}}{2},\qquad C_1^+=\dfrac{i \Gamma_{30}(\rho_{-+}^{ss}-\rho_{+-}^{ss})\rho_{11}^{ss}}{\Omega}\nonumber\\
\lambda_1^-=&\dfrac{2\Gamma_{20}+\Gamma_{21}}{2},\qquad C_1^-=0.\label{c24}
\end{align}
It is clear that the Lorentzian with ultra-narrow linewidth $\lambda_1^-$ vanished when $C_1^-=0$. As we analyzed before, the other narrow linewidth $\lambda_1^+$ has increased.
Whereas the accumulation on $\rho_{11}^{ss}$ and the decrease of  $\Gamma_{30}$ compete with each other, which makes the amplitude factor  $C_1^+$ basically unchanged, so that the height of the narrow peak  $C_1^+/\lambda_1^+$ is reduced. Therefore, in this case,  we can only get an inconspicuous narrow peak in Fig.~\ref{fig6}, but not an sharp ultranarrow peak in Fig.~\ref{fig4}.

In summary, for the case of energy level crossing that $g/\omega_q=0.7056$, the two states $\vert\psi_2\rangle$ and $\vert\psi_3\rangle$ are degenerate. When the photons are driven from the ground state to  $\vert\psi_3\rangle$, the population will not  be transferred to $\vert\psi_2\rangle$, so that the electron shelving only exists in $\vert\psi_1\rangle$, which makes the four-level system here equivalent to the three-level system in Ref.~\cite{PhysRevA.93.033801}. And the narrowing of the spectrum is not obvious. For the case of avoided-crossing that $g/\omega_q=0.2$, there are two metastable states $\vert\psi_1\rangle$ and $\vert\psi_2\rangle$ with electronic shelving, corresponding to two Lorentzians with narrow linewidths, one of which is an ultranarrow peak. Therefore, we claim that the emission spectrum with narrower linewidth can be obtained by adjusting the coupling strength of the qubit-cavity ultrastrong coupling system.

\section{Conclusion}\label{sec5}
In this paper, we have investigated the cavity emission spectrum of a double qubit-cavity coupling system in the regime of USC. Owing to the ultrastrong coupling between qubits and cavity, the RWA is invalid, and the counter-rotating wave terms  must be taken into account. In the energy spectrum of quantum Rabi Hamiltonian, we find level crossing between  the second and third energy levels, and  avoided-crossing between  the third and fourth energy levels. In order to further investigate the fluorescence spectra in these two cases, we fix the frequency of the driving field to resonate with the transition from the ground state to the third energy level. Hence the atomic system is reduced to a four-level dressed state model driven by a light field with Rabi frequency. Different from the Mollow triplet in the two-level systems, we find an ultranarrow peak at the line center. The narrow linewidth is derived from the  slowly decaying incoherent term $\rho_{++}+\rho_{--}$ and $\rho_{11}-\rho_{22}$ in the density matrix, and the  physical  origin is related to the quantum interference between two transition  pathways in the $\Lambda$-type , V-type, and $\Xi$-type three-level structure. Furthermore, in the case of level crossing, the population on $\rho_{11}$ grows as the coupling strength  increases, thus the  transitions from $\vert\psi_{1}\rangle$ to $\vert+\rangle$ and $\vert-\rangle$ become more significant. Hence, a clear inner sidebands can be found in the fluorescence spectrum. Our results indicate  that the spectral ultranarrowing can be achieved by adjusting the coupling strength of the USC system appropriately. And  the process of exploring the physical origin for the narrow peak can be helpful for the study of quantum interference effect in the USC system.

\begin{backmatter}
\bmsection{Funding}
This work is supported by the National Natural Science Foundation of China (Grant No. 11774118).

\bmsection{Disclosures} 
The authors declare no conflicts of interest.

\bmsection{Supplemental document} 
See Supplement 1 for supporting content.

\end{backmatter}

\bibliography{sample}

\begin{thebibliography}{10}
\newcommand{\enquote}[1]{``#1''}

\bibitem{RevModPhys.73.357}
Y.~Makhlin, G.~Sch\"on, and A.~Shnirman, \enquote{Quantum-state engineering
  with josephson-junction devices,} {\protect\JournalTitle{Rev. Mod. Phys.}}
  \textbf{73}, 357--400 (2001).

\bibitem{Vion886}
D.~Vion, A.~Aassime, A.~Cottet, P.~Joyez, H.~Pothier, C.~Urbina, D.~Esteve, and
  M.~H. Devoret, \enquote{Manipulating the quantum state of an electrical
  circuit,} {\protect\JournalTitle{Science}} \textbf{296}, 886--889 (2002).

\bibitem{doi:10.1063/1.5089550}
P.~Krantz, M.~Kjaergaard, F.~Yan, T.~P. Orlando, S.~Gustavsson, and W.~D.
  Oliver, \enquote{A quantum engineer's guide to superconducting qubits,}
  {\protect\JournalTitle{Appl. Phys. Rev.}} \textbf{6}, 021318 (2019).

\bibitem{Hennessy2007}
K.~Hennessy, A.~Badolato, M.~Winger, D.~Gerace, M.~Atat{\"u}re, S.~Gulde,
  S.~F{\"a}lt, E.~L. Hu, and A.~Imamo{\u{g}}lu, \enquote{Quantum nature of a
  strongly coupled single quantum dot--cavity system,}
  {\protect\JournalTitle{Nature (London)}} \textbf{445}, 896--899 (2007).

\bibitem{Wendin_2017}
G.~Wendin, \enquote{Quantum information processing with superconducting
  circuits: a review,} {\protect\JournalTitle{Rep. Prog. Phys.}} \textbf{80},
  106001 (2017).

\bibitem{Chiorescu1869}
I.~Chiorescu, Y.~Nakamura, C.~J. P.~M. Harmans, and J.~E. Mooij,
  \enquote{Coherent quantum dynamics of a superconducting flux qubit,}
  {\protect\JournalTitle{Science}} \textbf{299}, 1869--1871 (2003).

\bibitem{Fedorov2012}
A.~Fedorov, L.~Steffen, M.~Baur, M.~P. da~Silva, and A.~Wallraff,
  \enquote{Implementation of a toffoli gate with superconducting circuits,}
  {\protect\JournalTitle{Nature (London)}} \textbf{481}, 170--172 (2012).

\bibitem{Reed2012}
M.~D. Reed, L.~DiCarlo, S.~E. Nigg, L.~Sun, L.~Frunzio, S.~M. Girvin, and R.~J.
  Schoelkopf, \enquote{Realization of three-qubit quantum error correction with
  superconducting circuits,} {\protect\JournalTitle{Nature (London)}}
  \textbf{482}, 382--385 (2012).

\bibitem{Ofek2016}
N.~Ofek, A.~Petrenko, R.~Heeres, P.~Reinhold, Z.~Leghtas, B.~Vlastakis, Y.~Liu,
  L.~Frunzio, S.~M. Girvin, L.~Jiang, M.~Mirrahimi, M.~H. Devoret, and R.~J.
  Schoelkopf, \enquote{Extending the lifetime of a quantum bit with error
  correction in superconducting circuits,} {\protect\JournalTitle{Nature
  (London)}} \textbf{536}, 441--445 (2016).

\bibitem{Haroche2020}
S.~Haroche, M.~Brune, and J.~M. Raimond, \enquote{From cavity to circuit
  quantum electrodynamics,} {\protect\JournalTitle{Nat. Phys.}} \textbf{16},
  243--246 (2020).

\bibitem{Niemczyk2010}
T.~Niemczyk, F.~Deppe, H.~Huebl, E.~P. Menzel, F.~Hocke, M.~J. Schwarz, J.~J.
  Garcia-Ripoll, D.~Zueco, T.~H{\"u}mmer, E.~Solano, A.~Marx, and R.~Gross,
  \enquote{Circuit quantum electrodynamics in the ultrastrong-coupling regime,}
  {\protect\JournalTitle{Nat. Phys.}} \textbf{6}, 772--776 (2010).

\bibitem{PhysRevLett.105.237001}
P.~Forn-D\'{\i}az, J.~Lisenfeld, D.~Marcos, J.~J. Garc\'{\i}a-Ripoll,
  E.~Solano, C.~J. P.~M. Harmans, and J.~E. Mooij, \enquote{Observation of the
  bloch-siegert shift in a qubit-oscillator system in the ultrastrong coupling
  regime,} {\protect\JournalTitle{Phys. Rev. Lett.}} \textbf{105}, 237001
  (2010).

\bibitem{PhysRevLett.104.023601}
P.~Nataf and C.~Ciuti, \enquote{Vacuum degeneracy of a circuit qed system in
  the ultrastrong coupling regime,} {\protect\JournalTitle{Phys. Rev. Lett.}}
  \textbf{104}, 023601 (2010).

\bibitem{PhysRevLett.109.193602}
A.~Ridolfo, M.~Leib, S.~Savasta, and M.~J. Hartmann, \enquote{Photon blockade
  in the ultrastrong coupling regime,} {\protect\JournalTitle{Phys. Rev.
  Lett.}} \textbf{109}, 193602 (2012).

\bibitem{PhysRevLett.110.163601}
A.~Ridolfo, S.~Savasta, and M.~J. Hartmann, \enquote{Nonclassical radiation
  from thermal cavities in the ultrastrong coupling regime,}
  {\protect\JournalTitle{Phys. Rev. Lett.}} \textbf{110}, 163601 (2013).

\bibitem{PhysRevA.90.043817}
L.~Garziano, R.~Stassi, A.~Ridolfo, O.~Di~Stefano, and S.~Savasta,
  \enquote{Vacuum-induced symmetry breaking in a superconducting quantum
  circuit,} {\protect\JournalTitle{Phys. Rev. A}} \textbf{90}, 043817 (2014).

\bibitem{PhysRevA.92.063830}
L.~Garziano, R.~Stassi, V.~Macr\`{\i}, A.~F. Kockum, S.~Savasta, and F.~Nori,
  \enquote{Multiphoton quantum rabi oscillations in ultrastrong cavity qed,}
  {\protect\JournalTitle{Phys. Rev. A}} \textbf{92}, 063830 (2015).

\bibitem{PhysRevLett.117.043601}
L.~Garziano, V.~Macr\`{\i}, R.~Stassi, O.~Di~Stefano, F.~Nori, and S.~Savasta,
  \enquote{One photon can simultaneously excite two or more atoms,}
  {\protect\JournalTitle{Phys. Rev. Lett.}} \textbf{117}, 043601 (2016).

\bibitem{PhysRevA.95.063848}
P.~Zhao, X.~Tan, H.~Yu, S.~L. Zhu, and Y.~Yu, \enquote{Simultaneously exciting
  two atoms with photon-mediated raman interactions,}
  {\protect\JournalTitle{Phys. Rev. A}} \textbf{95}, 063848 (2017).

\bibitem{Kockum2017}
A.~F. Kockum, V.~Macr{\`i}, L.~Garziano, S.~Savasta, and F.~Nori,
  \enquote{Frequency conversion in ultrastrong cavity qed,}
  {\protect\JournalTitle{Sci. Rep.}} \textbf{7}, 5313 (2017).

\bibitem{PhysRevA.98.062327}
V.~Macr\`{\i}, F.~Nori, and A.~F. Kockum, \enquote{Simple preparation of bell
  and greenberger-horne-zeilinger states using ultrastrong-coupling circuit
  qed,} {\protect\JournalTitle{Phys. Rev. A}} \textbf{98}, 062327 (2018).

\bibitem{PhysRevA.101.053818}
V.~Macr\`{\i}, F.~Nori, S.~Savasta, and D.~Zueco, \enquote{Spin squeezing by
  one-photon--two-atom excitation processes in atomic ensembles,}
  {\protect\JournalTitle{Phys. Rev. A}} \textbf{101}, 053818 (2020).

\bibitem{oc283766}
S.~S. Shamailov, A.~S. Parkins, M.~J. Collett, and H.~J. Carmichael,
  \enquote{Multi-photon blockade and dressing of the dressed states,}
  {\protect\JournalTitle{Opt. Commun.}} \textbf{283}, 766 -- 772 (2010).

\bibitem{PhysRevA.91.043831}
W.~W. Deng, G.~X. Li, and H.~Qin, \enquote{Enhancement of the two-photon
  blockade in a strong-coupling qubit-cavity system,}
  {\protect\JournalTitle{Phys. Rev. A}} \textbf{91}, 043831 (2015).

\bibitem{PhysRevA.52.3333}
G.~C. Hegerfeldt and M.~B. Plenio, \enquote{Spectral structures induced by
  electron shelving,} {\protect\JournalTitle{Phys. Rev. A}} \textbf{52},
  3333--3343 (1995).

\bibitem{oc117560}
B.~Garraway, M.~Kim, and P.~Knight, \enquote{Quantum jumps, atomic shelving and
  monte carlo fluorescence spectra,} {\protect\JournalTitle{Optics
  Communications}} \textbf{117}, 560--569 (1995).

\bibitem{Kues2017}
M.~Kues, C.~Reimer, B.~Wetzel, P.~Roztocki, B.~E. Little, S.~T. Chu,
  T.~Hansson, E.~A. Viktorov, D.~J. Moss, and R.~Morandotti, \enquote{Passively
  mode-locked laser with an ultra-narrow spectral width,}
  {\protect\JournalTitle{Nat. Photonics}} \textbf{11}, 159--162 (2017).

\bibitem{Khivrich2020}
I.~Khivrich and S.~Ilani, \enquote{Atomic-like charge qubit in a carbon
  nanotube enabling electric and magnetic field nano-sensing,}
  {\protect\JournalTitle{Nat. Commun.}} \textbf{11}, 2299 (2020).

\bibitem{PhysRevA.42.1630}
L.~M. Narducci, M.~O. Scully, G.-L. Oppo, P.~Ru, and J.~R. Tredicce,
  \enquote{Spontaneous emission and absorption properties of a driven
  three-level system,} {\protect\JournalTitle{Phys. Rev. A}} \textbf{42},
  1630--1649 (1990).

\bibitem{PhysRevLett.66.2460}
D.~J. Gauthier, Y.~Zhu, and T.~W. Mossberg, \enquote{Observation of linewidth
  narrowing due to coherent stabilization of quantum fluctuations,}
  {\protect\JournalTitle{Phys. Rev. Lett.}} \textbf{66}, 2460--2463 (1991).

\bibitem{Vus}
T.~N. Vu, A.~Klehr, B.~Sumpf, H.~Wenzel, G.~Erbert, and G.~Tr\"{a}nkle,
  \enquote{Tunable 975\&\#x2009;\&\#x2009;nm nanosecond diode-laser-based
  master-oscillator power-amplifier system with 16.3\&\#x2009;\&\#x2009;w peak
  power and narrow spectral linewidth below 10\&\#x2009;\&\#x2009;pm,}
  {\protect\JournalTitle{Opt. Lett.}} \textbf{39}, 5138--5141 (2014).

\bibitem{Changs}
W.~K. Chang, H.~P. Chung, Y.~Y. Chou, R.~Geiss, S.~D. Yang, T.~Pertsch, and
  Y.~H. Chen, \enquote{Electro-optically spectrum narrowed, multiline
  intracavity optical parametric oscillators,} {\protect\JournalTitle{Opt.
  Express}} \textbf{24}, 28905--28914 (2016).

\bibitem{PhysRevLett.118.063601}
S.~Hughes and G.~S. Agarwal, \enquote{Anisotropy-induced quantum interference
  and population trapping between orthogonal quantum dot exciton states in
  semiconductor cavity systems,} {\protect\JournalTitle{Phys. Rev. Lett.}}
  \textbf{118}, 063601 (2017).

\bibitem{PhysRevX.5.021035}
N.~M. Sundaresan, Y.~Liu, D.~Sadri, L.~J. Sz\H{o}cs, D.~L. Underwood,
  M.~Malekakhlagh, H.~E. T\"ureci, and A.~A. Houck, \enquote{Beyond strong
  coupling in a multimode cavity,} {\protect\JournalTitle{Phys. Rev. X}}
  \textbf{5}, 021035 (2015).

\bibitem{PhysRevX.6.031004}
D.~M. Toyli, A.~W. Eddins, S.~Boutin, S.~Puri, D.~Hover, V.~Bolkhovsky, W.~D.
  Oliver, A.~Blais, and I.~Siddiqi, \enquote{Resonance fluorescence from an
  artificial atom in squeezed vacuum,} {\protect\JournalTitle{Phys. Rev. X}}
  \textbf{6}, 031004 (2016).

\bibitem{PhysRevA.93.033801}
H.~M. Castro-Beltr\'an, R.~Rom\'an-Ancheyta, and L.~Guti\'errez,
  \enquote{Phase-dependent fluctuations of intermittent resonance
  fluorescence,} {\protect\JournalTitle{Phys. Rev. A}} \textbf{93}, 033801
  (2016).

\bibitem{PhysRevA.92.023842}
K.~K.~W. Ma and C.~K. Law, \enquote{Three-photon resonance and adiabatic
  passage in the large-detuning rabi model,} {\protect\JournalTitle{Phys. Rev.
  A}} \textbf{92}, 023842 (2015).

\bibitem{PhysRevA.89.033827}
J.~F. Huang and C.~K. Law, \enquote{Photon emission via vacuum-dressed
  intermediate states under ultrastrong coupling,} {\protect\JournalTitle{Phys.
  Rev. A}} \textbf{89}, 033827 (2014).

\bibitem{PhysRevA.88.063829}
L.~Garziano, A.~Ridolfo, R.~Stassi, O.~Di~Stefano, and S.~Savasta,
  \enquote{Switching on and off of ultrastrong light-matter interaction: Photon
  statistics of quantum vacuum radiation,} {\protect\JournalTitle{Phys. Rev.
  A}} \textbf{88}, 063829 (2013).

\bibitem{PhysRevA.95.023829}
A.~Le~Boit\'e, M.-J. Hwang, and M.~B. Plenio, \enquote{Metastability in the
  driven-dissipative rabi model,} {\protect\JournalTitle{Phys. Rev. A}}
  \textbf{95}, 023829 (2017).

\bibitem{mollow}
B.~R. Mollow, \enquote{Power spectrum of light scattered by two-level systems,}
  {\protect\JournalTitle{Phys. Rev.}} \textbf{188}, 1969--1975 (1969).

\bibitem{PhysRevLett.77.3995}
P.~Zhou and S.~Swain, \enquote{Ultranarrow spectral lines via quantum
  interference,} {\protect\JournalTitle{Phys. Rev. Lett.}} \textbf{77},
  3995--3998 (1996).

\bibitem{PhysRevA.56.3011}
P.~Zhou and S.~Swain, \enquote{Quantum interference in resonance fluorescence
  for a driven v atom,} {\protect\JournalTitle{Phys. Rev. A}} \textbf{56},
  3011--3021 (1997).

\end{thebibliography}

\end{document}